\def\etal{{et al.\thinspace}}

\def\eg{{\em e.g.\ }}

\def\spose#1{\hbox to 0pt{#1\hss}}
\def\approxlt{\mathrel{\spose{\lower 3pt\hbox{$\sim$}}
        \raise 2.0pt\hbox{$$<$$}}}
\def\approxgt{\mathrel{\spose{\lower 3pt\hbox{$\sim$}}
        \raise 2.0pt\hbox{$>$}}}

\def\multleft#1{\hbox to size{\vbox {\halign {\lft{##}\cr #1}}\hfill}\par}
\def\multright#1{\hbox to size{\vbox {\halign {\rt{##}\cr #1}}\hfill}\par}

\def\today{\ifcase\month\or January\or February\or March\or April\or May\or
      June\or July\or August\or September\or October\or November\or December\fi
      \space\number\day, \number\year}
\def\s{\hbox{\phantom{5}}}      

\def\boxit#1{\vbox{\hrule\hbox{\vrule\kern3pt\vbox{\kern3pt
          #1 \kern3pt}\kern3pt\vrule}\hrule}}

\def\km{{\rm\thinspace km}}

\def\Lsun{\hbox{$\rm\thinspace L_{\odot}$}}

\def\Mpc{{\rm\thinspace Mpc}}
\def\Msun{\hbox{$\rm\thinspace M_{\odot}$}}

\def\s{{\rm\thinspace s}}

\def\kmps{\hbox{$\km\s^{-1}\,$}}

\def\nobs{57}
\def\ndet{9}
\def\nstat{53}

\documentstyle[epsfig,psfig]{mn}
\begin{document}
\hsize=6truein

\title[Submillimetre properties of $z\sim2$ quasars]
{The SCUBA Bright Quasar Survey II: unveiling the quasar epoch
at submillimetre wavelengths}
\author[R.S. Priddey et al.]
{\parbox[]{6.5in}
{Robert S. Priddey$^1$, Kate G. Isaak$^2$, Richard G. McMahon$^3$ \&
Alain Omont$^4$}
\\ 
$^1${\it Astrophysics Group, Imperial College, Blackett Laboratory,
Prince Consort Road, London SW7 2BZ}\\
$^2${\it Cavendish Astrophysics, University of Cambridge, Cambridge CB3 0HE}\\
$^3${\it Institute of Astronomy, Madingley Road, Cambridge CB3 0HA, UK}\\
$^4${\it Institut d'Astrophysique de Paris, CNRS, 98bis Bd. Arago, Paris, 
France}\\
email: r.priddey@ic.ac.uk, isaak@mrao.cam.ac.uk, rgm@ast.cam.ac.uk}

\date{Accepted for publication in MNRAS, 14th November 2002}

\maketitle

\begin{abstract}
We present results of the first systematic search for submillimetre
(submm) continuum emission from $z\sim2$, radio-quiet,
optically-luminous ($M_B<-27.5$) quasars, using the SCUBA array camera on
the JCMT.
We have observed a homogeneous sample of \nobs\ quasars in the
redshift range $1.5<z<3.0$--- the epoch during which the comoving 
density of luminous
AGN peaks--- to make a systematic comparison
with an equivalent sample at high redshift
($z>4$; Isaak et al., 2002: Paper I).
The target sensitivity of the survey, $3\sigma=10$mJy at 850$\mu$m,
was chosen to enable efficient identification of bright submm 
sources, suitable for detailed follow-up.
\ndet\ targets are detected with $3\sigma$ significance or greater,
with fluxes in the range 7--17mJy.
Although the detection rate above 10mJy is lower than that of the
$z>4$ survey, the weighted mean flux of the undetected sources,
1.9$\pm$0.4mJy, is similar to that at $z>4$ (2.0$\pm$0.6mJy).
The statistical significance of trends is analysed, and it is found
that: (i) within the limited optical luminosity range studied, there
is no strong evidence for a correlation between submm and optical
luminosity; (ii) there is a suggestion of a variation of
submm detectability with redshift, but that this is consistent with
the $K$-correction of a characteristic far-infrared spectrum.

\end{abstract}



\section{INTRODUCTION}
By virtue of their high, sustained luminosity across the spectrum,
quasars are ideal targets for studies of the evolution of structure
throughout the history of the cosmos.
Whilst a rare phenomenon in our local volume of space,
the comoving density of luminous AGN rises sharply as one goes back in time,
to a peak (Schmidt, Schneider \& Gunn., 1995; Fan et al. 2001)
or a plateau (Miyaji, Hasinger \& Schmidt, 2000) around $z$=2, an
epoch dependence echoing the star formation history of galaxies (e.g. 
Madau et al. 1996; Steidel et al. 1999). 
It is widely believed that most massive galaxies undergo a short-lived AGN 
phase, as they build up the supermassive black holes observed in local 
galactic cores (Magorrian et al., 1998).
Thus in observing a quasar, one catches its host galaxy at a
significant period in its evolution.

At far-infrared (FIR) to millimetre (mm) wavelengths, AGN have long
been recognised as amongst the most prominent high-redshift sources
(e.g. the compilations in McMahon et al., 1999 (M99); Rowan-Robinson, 2000).
While there is strong evidence that emission from this region of the 
spectrum is thermal reradiation from warm dust (e.g. Hughes et al., 1993), 
it is not always clear that the AGN itself is the sole contributor
to dust heating, and it is plausible that a fraction of the submm
luminosity could derive from stars in the surrounding galaxy.
There would thus be a direct overlap between dusty quasars and
SCUBA survey sources, which, it is argued, are the star-forming
progenitors of massive spheroids.
Hence, targetted submm and mm surveys of high-redshift
AGN are valuable in elucidating the nature of cosmological 
submm sources,
and the relation between the evolution of the host galaxy and
its central black hole.

M99 observed a small sample of $z>4$, radio-quiet 
quasars with
the JCMT's SCUBA submm camera to high sensitivity 
($\sigma_{850\mu\rm m}\approx1.5$mJy), 
to determine
the submm properties of the ``typical'' high-redshift AGN.
A complementary, broad-but-shallow 
($\sigma_{850\mu\rm m}\approx3$mJy)
strategy was adopted by 
Isaak et al. (2002: I02)---
the SCUBA Bright Quasar Survey (SBQS) (Paper I)--- with the principal aim
of defining a statistically-significant sample of submm sources bright 
enough to permit a range of follow-up study.
Roughly a quarter of the targets were brighter than the $3\sigma\sim10$mJy
limit, suggesting that
a substantial fraction of high-redshift quasars 
have far-infrared luminosities comparable to their blue luminosities. 
A natural question is whether this ubiquitous 
submm activity
discovered at $z>4$ is typical of high-redshift
AGN in general: is the optically-luminous quasar phase always
accompanied by a dust-rich submillimetre source, 
or does this only occur at the highest redshifts?
In the current paper we present the results of a comparative survey
designed to address this question,
targetting the ``AGN epoch'' at $z\sim2$, 
the era at which the space density of quasars
reaches its maximum, and by which most ($>80$ percent)
of the matter that will ever be accreted onto supermassive
black holes has already served as fuel for AGN.
A presentation of these results, along with brief analysis,
is the purpose of the current paper: a more detailed and wide-ranging
study, in the context of all the recent mm and submm quasar surveys 
in this series (M99; I02; Omont et al., 2001; Omont et al., 2002),
will be given in a forthcoming work (Priddey et al., in prep.). 

We assume the currently-favoured $\Lambda$-dominated cosmology
$\Omega_M$=0.3, $\Omega_{\Lambda}$=0.7, $H_0$=65\kmps\Mpc$^{-1}$
($\Lambda$).
For continuity with previous work, we will give alternatives in 
an Einstein--de Sitter cosmology, $\Omega_M$=1, $\Omega_{\Lambda}$=0,
with $H_0$=50\kmps\Mpc$^{-1}$ (EdS).

\begin{figure}
\psfig{figure=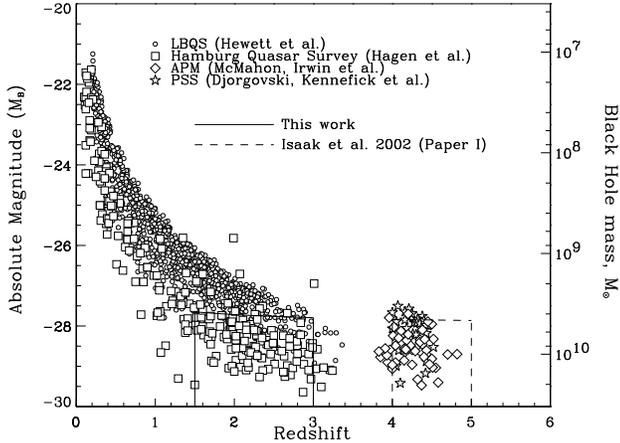,width=90mm}
\label{fig:hubble_diagram}
\caption{Hubble diagram for quasar surveys used in the Paper I (Isaak et al.
2002, $z>4$ selection) and the present work ($z\approx2$ selection).
Note that in each case, the optical luminosity was designed to be 
$M_B<-27.5$ {\it in the EdS cosmology} for consistency with
previous work. 
This is converted to the $\Lambda$ cosmology employed in this paper,
but it is clear that the difference between the samples is small.
The black hole mass (right-hand axis) is calculated
from $M_B$ assuming accretion at the Eddington rate.}
\end{figure}

\section{SAMPLE SELECTION AND OBSERVATIONS}
Our aim was to find a sample of bright, medium-redshift quasars
well-matched to the $z>4$ sample observed by Isaak et al. (2002) 
(Figure 1).
To avoid too heterogeneous a sample, we restricted the input
catalogues to a few large, homogeneous surveys.
The final target list comprises quasars preselected from the 
Large Bright Quasar Survey (LBQS: Hewett, Foltz \& Chaffee, 1995) and the
Hamburg Quasar Survey (HS: Engels et al. 1998; Hagen et al. 1999).
The first selection criterion is based on optical luminosity 
represented by absolute $B$-band magnitude ($M_B$).
At $z=2$, the $J$ photometric band samples very close to rest-frame $B$,
and $H$ starts to do so towards higher redshifts. 
Using a combination of these bands 
therefore gives an absorption-free 
assessment of the rest-frame optical magnitude,
minimizing the error due extrapolation of the continuum.
This overcomes many of the problems encountered when deriving $M_B$ for
$z>4$ quasars--- e.g. contamination of the $R$ band by strong
emission and absorption features (as detailed in Paper I).
Similarly, at $z\approx2$, observed-frame $B$ becomes compromised by the
strong CIV and Ly-$\alpha$ emission lines. 

Hence, to obtain a sample of luminous $z\approx2$ quasars all with $J$ and
$H$ magnitudes, 
we cross-correlated catalogues of bright, medium-redshift quasars 
with the 2MASS near-infrared catalogue through the web-based interface at
IPAC (http://www.ipac.caltech.edu). 
Counting sources out to a radius of 60 arcsec enabled us
to determine a maximum association radius of 5 arcsec,
within which the probability of a chance association is $<$0.5 percent.
Absolute $B$-band magnitudes were
calculated from apparent $J$ and $H$ magnitudes thereby obtained
(all our targets were detected in both bands)
weighted according to the
proximity of the band to $B$ in the redshifted spectrum.
They have been corrected for Galactic extinction, though in most cases
this is negligible ($<0.05$mag).
We have assumed an optical spectral index $\alpha_{\rm opt}=-0.5$ (where
$f_{\nu}\propto\nu^{\alpha}$), though, as noted, the extrapolation error
is small. Photometric errors in $J$ and $H$ are typically 0.1--0.2mag.

In order to match optical luminosity with the $z>4$ sample, 
objects with $M_B^{\rm EdS}<-27.5$ were initially selected--- 
i.e. adopting the same
cosmology as per the selection of the $z>4$ 
sample\footnote{Note that this selection was made in the EdS cosmology;
as shown in Figure 1 and Table 1, converting to the $\Lambda$ cosmology 
introduces an average shift of $-$0.2mag. Most importantly, 
the overall difference between the $z=2$ and $z>4$ samples is small.}. 
A wide redshift range
$1.50<z<3.00$ was chosen to maximize the potential number of targets
(Figure 1).
To derive an observing sequence, the final selection was prioritized 
according to optical luminosity, the most luminous given preference.

The nominal completeness limit for 2MASS is 15.8 and 15.1 in $J$ and $H$ 
respectively. However, at the high Galactic latitudes inhabited
by these quasars, accurate detections up to one magnitude better than this 
can be achieved (Cutri et al., 2000).
From this and our extensive quasar catalogue 
cross-correlations with 2MASS, we estimate that at $z=3$, the 2MASS limit
corresponds to $M_B^{\rm EdS}\approx-27.6$, thus for the greater
extent of our redshift range we are not in danger of overlooking bright
potential targets that failed to be detected by 2MASS.

The target list was correlated with the NRAO VLA Sky Survey (NVSS)
to identify radio-loud sources, which were excised from the
list. As discussed in M99 and I02, ``radio loud'' in this context 
primarily refers to the extent of contamination of the thermal submm 
continuum by the extrapolated radio synchrotron. Assuming a spectral index
$\alpha=-0.5$, a source with $S_{\rm 1.4GHz}<1.5$mJy (below the NVSS limit)
will have $S_{\rm850\mu m}<0.1$mJy.

The JCMT observing strategy for this project was 
the same as for the
comparable $z>4$ program described in detail in I02.
To reiterate: the required RMS (1$\sigma=3$mJy) can be reached in 
a short amount of
observing in relatively poor (zenith transmission 65\%)
weather, so the JCMT fallback queue was utilized.
SCUBA was employed in photometry mode, that is, the source is placed
on the central bolometer, while the rest of the array samples the sky.
Flux calibration was determined from the planets Mars
and Uranus, or from secondary continuum standards (CRL618, OH231.8, IRC10216).
The APM catalogue
(online interface http://www.ast.cam.ac.uk/$\sim$apmcat)
was used to verify coordinates, which are given in Table 1.

\begin{table*}
\caption{APM astrometry, and 2MASS NIR and SCUBA submm
photometry of $z\sim2$ quasars.}
\label{tab:z2-phot}
\begin{tabular}{lllcccccrl}
Target name & RA & Dec. & $z$ & $J$ & $H$ & $M_B^{\Lambda}$ ($M_B^{\rm EdS}$) 
& Observation & $S_{850}\pm\sigma_{850}$ & Notes\\
&(J2000)&(J2000)&&&&&Dates&(mJy)&\\
(1) & (2) & (3) & (4) & (5) & (6) & (7) & (8) & (9) & (10)\\
\hline
LBQS B0018$-$0220$^a$ & 00 21 27.37 & $-$02 03 33.8 & 
2.56 & 16.2 & 15.4 & $-$28.6($-$28.3) & 05/08/00& 17.2$\pm$2.9&\\
HS B0035+4405$^a$     & 00 37 52.31 & +44 21 32.9   &2.71 & 15.9 & 15.4 & $-$28.7($-$28.4) & 22/09/00& 9.4$\pm$2.8\\
HS B0211+1858         & 02 14 29.71 & +19 12 37.6   &2.47 & 16.2 & 15.5 & $-$28.3($-$28.0) & 22/09/00& 7.1$\pm$2.1& \\
HS B0810+2554$^b$     & 08 13 31.30 & +25 45 02.9   &1.50 & 14.1 & 13.2 & $-$29.2($-$29.1) & 07,09,10/08/00&7.6$\pm$1.8&\\
HS B0943+3155         & 09 46 23.21 & +31 41 30.5 & 2.79 & 16.5 & 16.0 & $-$28.1($-$27.8) & 08/10/00& 9.6$\pm$3.0& \\
HS B1140+2711         & 11 42 54.27 & +26 54 57.8 & 2.63 & 15.8 & 15.2 & $-$28.8($-$28.5) & 17/01/01& 8.6$\pm$2.6&\\
HS B1141+4201         & 11 43 52.04 & +41 45 19.8 & 2.12 & 15.5 & 15.1 & $-$28.5($-$28.3) & 18/01/01& 8.6$\pm$2.6&BAL\\
HS B1310+4308         & 13 12 48.73 & +42 52 36.8 & 2.60 & 16.0 & 15.8 & $-$28.2($-$27.9) & 13/04/01& 10.0$\pm$2.8&weak BAL\\
HS B1337+2123         & 13 40 10.84 & +21 08 44.5* & 2.70 & 16.5 & 15.6 & $-$28.4($-$28.1) & 13/04/01& 6.8$\pm$2.1&\\
\hline
LBQS B0009+0219       & 00 12 19.64 & +02 36 35.4 &2.64 & 16.3 & 16.0 
& $-$28.0($-$27.7) & 06/08/00& 1.4$\pm$3.2&\\
LBQS B0009$-$0138     & 00 12 10.91 &$-$01 22 07.7&2.00 & 16.2 & 15.7 & $-$27.6($-$27.4) & 06/08/00& 2.2$\pm$2.8&\\
LBQS B0013$-$0029     & 00 16 02.41 &$-$00 12 25.2&2.08 & 16.3 & 15.2 & $-$27.7($-$27.5) & 06/08/00& 3.2$\pm$3.3&\\
HS B0017+2116	      & 00 20 10.85* &+21 32 51.4*&2.02 & 16.0 & 15.5& $-$27.8($-$27.6) & 22/09/00& 2.3$\pm$3.0&\\
LBQS B0025$-$0151     & 00 27 33.82 &$-$01 34 52.4&2.08 & 16.3 & 15.7 & $-$27.7($-$27.5) & 06/08/00& 4.8$\pm$2.4&BAL\\
HS B0029+3725$^c$     & 00 32 10.08 & +37 42 32.5 &1.85 & 15.8 & 15.5 & $-$28.6($-$28.4) & 14/09/00& 2.2$\pm$2.6&\\ 
HS B0036+3842         & 00 39 07.49  &+38 59 15.5*&2.36 & 16.4 & 15.6 & $-$28.0($-$27.7) & 22/09/00& 3.0$\pm$3.0&BAL\\
HS B0037+1351         & 00 40 23.76  & +14 08 07.5&1.87 & 15.6 & 15.1 & $-$28.1($-$27.9) & 06/08/00& 3.4$\pm$3.1&\\
HS B0042+3704         & 00 44 48.3$^{\dagger}$  & +37 21 14$^{\dagger}$ &2.41 & 16.4 & 15.9 
& $-$28.0($-$27.7) & 12/10/00& $-$1.2$\pm$2.6&\\
HS B0105+1619         & 01 08 06.47  & +16 35 50.4&2.64 & 15.7 & 15.1 
& $-$28.9($-$28.6) & 04/08/00& 3.8$\pm$2.6&\\
HS B0119+1432         & 01 21 56.06  & +14 48 24.0&2.87 & 15.5 & 15.1 & $-$29.0($-$28.7) & 04/08/00& 3.6$\pm$2.6&\\
HS B0150+3806         & 01 53 13.55  & +38 21 24.2&1.96 & 16.0 & 15.4 & $-$27.8($-$27.6) & 06/08/00& 2.2$\pm$3.0&\\
HS B0202+1848         & 02 05 27.52  & +19 02 29.8&2.70 & 15.7 & 15.3& $-$28.8($-$28.5) & 05/08/00& 0.6$\pm$3.3& \\
HS B0218+3707    &  02 21 05.52   &   +37 20 46.2 &2.41 & 15.8 & 14.9& $-$28.6($-$28.4) & 04/08/00& 1.9$\pm$2.6&\\
HS B0219+1452    &  02 22 31.71*   &  +15 06 28.6*&1.71 & 14.8 & 14.2 & $-$28.8($-$28.6) & 05/08/00& $-$5.3$\pm$2.8&\\
HS B0248+3402    &  02 51 27.78   &   +34 14 42.1 &2.23 & 15.4 & 15.0 & $-$28.6($-$28.4) & 04/08/00& $-$0.2$\pm$2.7&\\
HS B0752+3429    &  07 55 24.10   &   +34 21 34.2 &2.11 & 16.1 & 15.4 & $-$27.9($-$27.7) & 22/09/00& 2.4$\pm$2.8 &\\  
HS B0800+3031    &  08 03 42.05   &   +30 22 54.8 &2.02 & 15.1 & 14.5 & $-$28.8($-$28.6) & 07/07/00& $-$3.0$\pm$3.4 &\\ 
HS B0808+1218    &  08 10 56.9$^{\dagger}$ & +12 09 14$^{\dagger}$ &2.26 
& 16.2 & 15.5 & $-$28.0($-$27.8) & 22/09,08/10/00& 1.1$\pm$2.1 & \\  
HS B0821+3613    &  08 25 07.66   &   +36 04 11.5 &1.58 & 15.5 & 14.7 
& $-$27.9($-$27.7) & 22/09,08/10/00 & 1.5$\pm$2.0 &\\ 
HS B0830+1833    &  08 32 55.63   &   +18 23 00.7 &2.27 & 15.9 & 15.4 
& $-$28.3($-$28.1) & 08/10/00& $-$0.5$\pm$3.1 & \\ 
HS B0834+1509    &  08 37 12.87*   &   +14 59 17.5&2.51 & 16.2 & 15.4 & $-$28.3($-$28.0) & 08/10/00& $-$1.0$\pm$2.8 & \\ 
HS B0926+3608    &  09 29 52.14   &   +35 54 49.8 &2.14 & 16.3 & 15.5 & $-$27.7($-$27.5) & 08/10/00& 1.2$\pm$2.8 & \\  
HS B0929+3156    &  09 32 08.77   &   +31 43 28.0*&2.08 & 16.1 & 15.6 & $-$27.9($-$27.7) & 08/10/00& 5.7$\pm$3.0 &weak BAL\\  
HS B0931+2258    &  09 34 42.26   &   +22 44 39.5 &1.74 & 15.4 & 14.8 & $-$28.2($-$28.0) & 08/10/00& $-$3.5$\pm$2.8 & \\ 
HS B1002+4400    &  10 05 17.47   &   +43 46 09.3 & 2.08 & 15.5 & 15.0& $-$28.5($-$28.3) & 13,19/04/01& 8.2$\pm$2.9&\\  
HS B1031+1831    &  10 34 28.89   &   +18 15 32.4 &1.53 & 15.1 & 14.3 & $-$28.3($-$28.1) & 19/04/01& $-$0.1$\pm$2.7 & BAL\\ 
HS B1049+4033    &  10 51 58.71*   &  +40 17 37.0*&2.15 & 15.7 & 15.1 & $-$28.4($-$28.2) & 19/04/01& 3.9$\pm$3.2 & \\  
HS B1111+4033    &  11 13 50.94*   &   +40 17 21.5&2.18 & 16.0 & 15.7 & $-$28.1($-$27.9) & 19/04/01&  2.2$\pm$2.8 & \\  
HS B1115+2015    &  11 18 00.52*   &  +19 58 53.8*&1.93 & 15.5 & 15.0 & $-$28.4($-$28.2) & 19/04/01& 5.3$\pm$2.8 & BAL\\  
HS B1126+3639    &  11 28 57.84*   &  +36 22 50.3*&2.89 & 16.3& 15.8 & $-$28.3($-$28.0) & 19/04/01& $-$2.5$\pm$2.8 & \\ 
HS B1155+2640    &  11 57 41.91  &   +26 23 56.2  &2.80 & 16.6 & 15.8 & $-$28.3($-$28.0) & 19/04/01& 3.5$\pm$2.7 & \\    
HS B1200+1539    &  12 03 31.28* &   +15 22 54.4* &2.97 & 15.8 & 15.2 & $-$28.9($-$28.6) & 08/01/01& $-$5.1$\pm$3.1 & \\ 
LBQS B1210+1731$^d$  & 12 13 03.02 & +17 14 23.4* & 2.54 & 16.3 & 15.6 & $-$28.3($-$28.0) & 08/01/01& 1.3$\pm$2.8 & \\
HS B1215+2430    &  12 18 10.98   &   +24 14 10.8 &2.36 & 15.8 & 15.1 & $-$28.6($-$28.3) & 19/04/01& 6.0$\pm$2.7 & \\    
HS B1302+4226    &  13 04 25.56   &   +42 10 09.7 &1.91 & 15.3 & 14.7 & $-$28.5($-$28.3) & 18/01/01& 2.4$\pm$2.4 & weak BAL\\  
HS B1326+3923    &  13 28 23.73*   &   +39 08 17.8&2.32 & 15.4 & 14.8 & $-$28.8($-$28.6) & 08/01/01& 7.4$\pm$3.0 & \\  
LBQS B1334$-$0033  &  13 36 47.16  &$-$00 48 57.4 &2.80 & 16.3 & 15.6 & $-$28.0($-$27.8) & 08/01/01& 2.5$\pm$2.6 &\\  
HS B1356+3113    &  13 59 08.39   &   +30 58 30.8 &2.26 & 16.2 & 15.7 & $-$28.6($-$28.3) & 20/03/01& 3.2$\pm$3.0 & \\  
HS B1417+4722    &  14 19 51.84   &   +47 09 01.1 &2.27 & 15.7 & 15.1 & $-$28.5($-$28.3)& 18/01/01& 8.8$\pm$3.4&\\  
HS B1422+4224    &  14 24 35.96   &   +42 10 30.6*&2.21 & 16.2 & 16.0 & $-$27.7($-$27.5)& 19/04/01& 10.7$\pm$4.7 & \\   
HS B1703+5350    &  17 04 06.74   &   +53 46 53.6 &2.37 & 15.9 & 15.3 & $-$28.4($-$28.1) & 20/03/01& $-$0.1$\pm$2.6 &\\ 
HS B1754+3818    &  17 56 39.6$^{\dagger}$   &   +38 17 52$^{\dagger}$ &2.16 & 16.0 & 15.4 & $-$28.1($-$27.9)& 20/03/01& 0.7$\pm$2.5 & \\   
HS B2134+1531    &  21 36 23.86*   &   +15 45 08.4&2.13 & 15.6 & 14.8 & $-$28.4($-$28.2)& 04/08/00& $-$0.2$\pm$2.9 & \\  
LBQS B2244$-$0105  &  22 46 49.30  &$-$00 49 53.9 &2.03 & 15.9 & 15.7 & $-$27.9($-$27.7)& 14/09/00& 0.2$\pm$2.8& \\   
HS B2245+2531    &  22 47 27.40   &   +25 47 30.5 &2.15 & 15.4 & 14.9 & $-$28.5($-$28.3)& 04/08/00& 3.5$\pm$2.6 & \\   
HS B2251+2941    &  22 53 38.59   &   +29 57 12.5 &1.57 & 15.5 & 14.7 & $-$27.9($-$27.7)& 14/09/00& $-$1.2$\pm$2.6&\\ 
HS B2337+1845    &  23 39 44.77*   &  +19 01 51.1 &2.62 & 15.6 & 14.9 & $-$29.1($-$28.8)& 04/08/00& 2.4$\pm$4.0&\\  
									       
\hline		 		 		   			       
\hline									       
\end{tabular}								       
\begin{minipage}{170mm}
a. Detected at 450$\mu$m (Isaak et al., in prep.);
b. {\it IRAS} source, $S_{60}=284\pm40$mJy, $S_{100}=538\pm160$mJy.
Recent {\it HST} STIS imaging shows this quasar to be quadrupally lensed
(Reimers et al. 2002);
c. $S_{\rm1.4 GHz}=2.6$mJy; d. $S_{\rm1.4 GHz}=2.0$mJy\\
$^*$ APM position offset from published coordinates by 1 arcsec or more:
2MASS position favours APM\\
$^{\dagger}$ blending in APM, published (HS) coordinates used
\end{minipage}
\end{table*}

\section{RESULTS AND COMMENTARY}
A total of \nobs\ $z\sim2$ quasars have been observed with SCUBA in the
current survey, \ndet\ of which are detected 
with $>$3$\sigma$ significance at 850$\mu$m.
The median RMS flux is $\sigma_{850}=2.8$mJy,
with a narrow (0.4mJy) interquartile range.
Four of the sources have $3\sigma>10$mJy: in the following analysis,
we exclude these from our complete statistical sample, which thus
consists of \nstat\ sources.

\begin{figure}
\psfig{figure=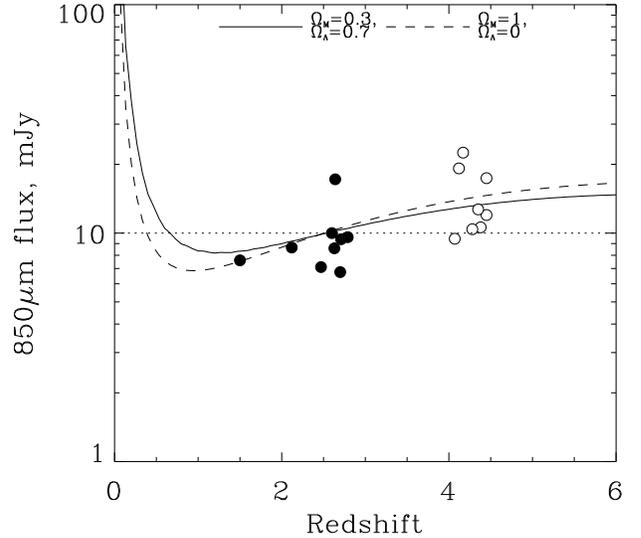,width=90mm}
\caption{Flux at 850$\mu$m against redshift, plotted for SBQS detections
from Paper I (I02, open symbols) and the current work 
(filled symbols). The dotted line represents the nominal 3$\sigma$=10mJy
limit for this work,
the approximate threshold for viable follow-up such as 
CO line detection.
The two curves are, for two different cosmologies, the flux one would
detect from a source of fixed luminosity having the mean isothermal SED
of PM01 ($T=40$K, $\beta=2$). They are arbitrarily normalized, and
are plotted only to illustrate the necessary $K$-correction between 
the redshift ranges of the two surveys.}
\label{fig:s850_z}
\end{figure}

One detection, the $z=2.6$ quasar LBQS B0018$-$0220,
is significantly brighter than the nominal 10mJy limit: 
with $S_{850}=17\pm3$mJy,
this is an exceptionally
bright source, comparable to some of the brightest high-redshift
submm sources known.
To recover most of the other detections, 
one must go down to $\approx8$mJy.
In contrast, seven of the eight detections from the $z>4$ SCUBA Bright 
survey (I02) lie above 10mJy. In Figure \ref{fig:s850_z}, 
850$\mu$m fluxes of the $z\sim2$ sources reported in this paper
are plotted along with the $z>4$ sample described in I02.
The curves represent the flux one would observe from an object of 
fixed luminosity as a function of redshift, assuming it has the
mean far-infrared SED determined from $z>4$ quasars by 
Priddey \& McMahon (2001: PM01), characterized by an isothermal 
temperature and emissivity index $T=40$K and $\beta=2$, respectively.
The large $K$-correction towards high redshift
is one possible reason why the $z>4$ sources appear systematically brighter,
however it is not obvious that the same SED is valid
for all objects at all redshifts. 
(Indeed, one of the motivations for {\it this} project was to yield a sample
bright enough to be studied at a number of submm wavelengths, 
so that the typical FIR SED of a $z=2$ quasar could be determined.)

\section{STATISTICAL ANALYSIS}

\subsection{Submillimetre versus optical}
\begin{figure}
\psfig{figure=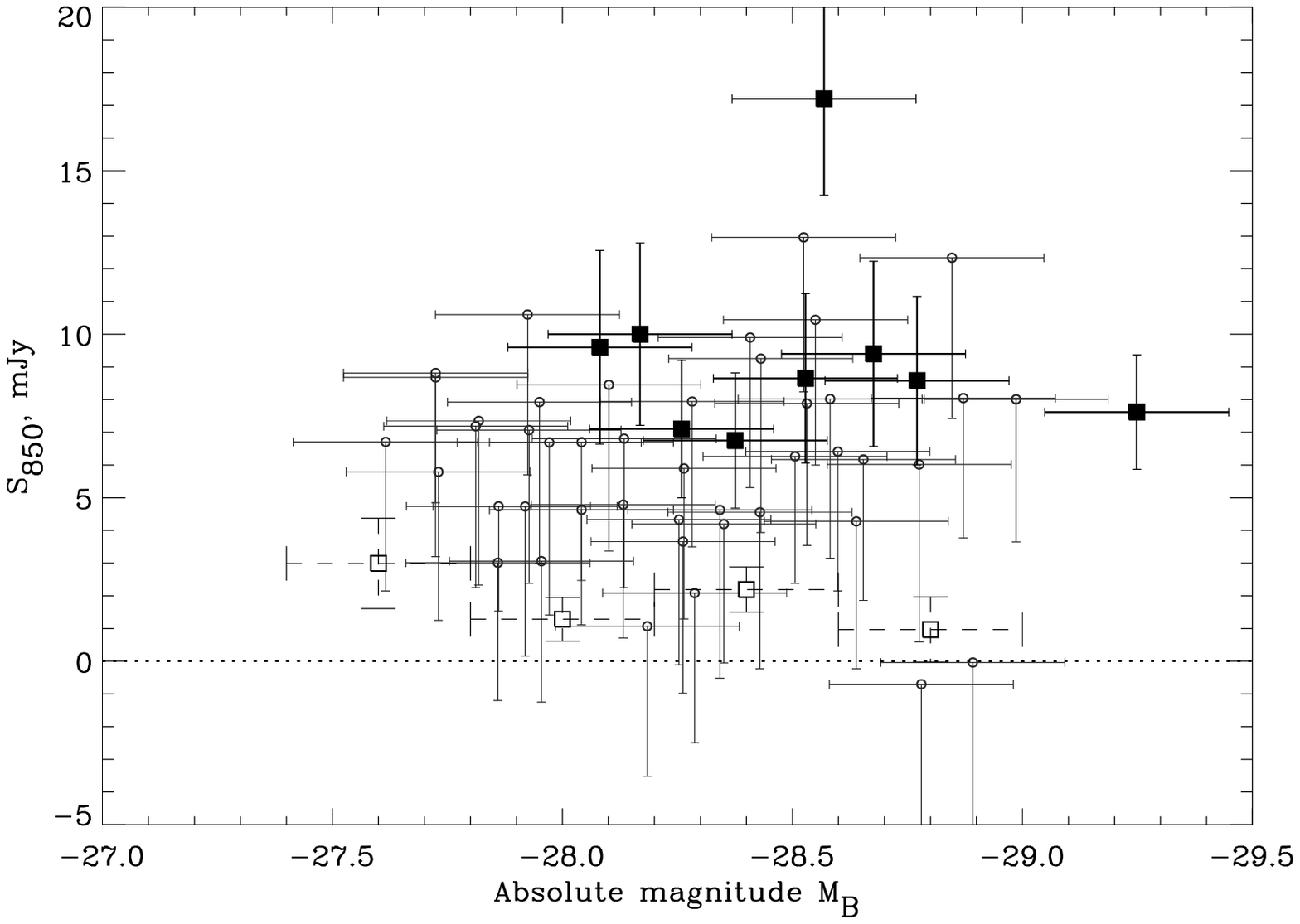,width=90mm}
\caption{850$\mu$m flux against absolute $B$ magnitude
($\Lambda$ cosmology) for the complete $z\approx2$ SCUBA sample. 
Submm detections
are plotted as large solid squares with error bars reflecting photometric
errors in $S_{850}$ and NIR magnitudes. 
Non-detections are plotted in light grey, as upper limits
corresponding to signal + 90\% confidence, with a vertical error bar
terminating at the value of the signal.
The weighted means of the non-detections in each of four magnitude bins
are also plotted as open symbols.}
\label{fig:s850mbz2}
\end{figure}

Figure \ref{fig:s850mbz2} shows 850$\mu$m fluxes plotted against 
absolute magnitudes for the $z\approx2$ sample. (NB: we do not correct
the optical for intrinsic absorption.)
Any correlation between submm and optical
is even less evident than
in the case of the $z>4$ sources of I02. This can be demonstrated
formally using statistical tests.
If there were a correlation between detectability and absolute magnitude,
we might expect the detected sources to be distributed with
respect to $M_B$ in a different way than the whole sample.
Applying the Kolmogorov--Smirnov (K--S) test informs us that 
the null hypothesis that the distributions are the same
Note that, unlike the $z>4$ objects, this material is less prone to
systematic error in the determination of $M_B$: the magnitudes were 
calculated homogeneously, and the need for continuum extrapolation was 
minimized.
Yet, the scatter remains, which suggests that it derives from
an underlying variance in the submm--optical relation.

As discussed in I02, one might naively 
expect a submm--optical correlation, whether 
the dust-heating source is stars (because spheroid mass scales with
black hole mass) or the AGN (because both $L_{\rm FIR}$ and $L_B$ scale
with bolometric luminosity). 
There is much scope for complexity which would smear out any correlation, 
for example the relative timing between AGN fuelling and starburst, 
or, in an AGN-powered scenario, the effects of varying the dust-torus geometry.
As a {\it caveat}, note that the nominal detection threshold, 10mJy,
corresponds, at $z=2$, to $L_{\rm FIR}\approx2\times10^{13}\Lsun\approx\nu L_B$
for the value of the magnitude cut, $M_B=-27.5$.
We are therefore probably sampling only the bright tail of the FIR
luminosity distribution, and we are biased towards sources lying just above 
this $L_{\rm FIR}\approx\nu L_B$ threshold. 
Thus for the detected sources, any such relation should be derived with 
caution.

\subsection{Optical colour}
As described in Section 2, 2MASS $J$ and $H$ magnitudes were obtained 
primarily to provide a homogeneous optical luminosity scale, directly
sampling rest-frame $B$.
However, we can use them to assess, roughly,
whether there is any correlation between submm detectability and
optical colour--- for example, whether submm excess is connected with
intrinsic reddening. 
The mean $B-H$ colour of the sample is 1.9$\pm$0.5,
whereas that of the detections is 2.0$\pm$0.5; similarly, $B-J$
of the sample is 1.3$\pm$0.4, of the detections 1.4$\pm$0.5. 
There is no evidence, then, that the detected quasars are 
redder than the underlying sample. However, given the large uncertainties,
further work is required to address the question systematically.

Note that while we are requiring our sample to be bright in rest-frame $B$,
the selection procedure for these surveys (e.g. LBQS $B_J<18.75$)
also requires these objects to be bright in the rest-frame UV.
This limit in observed $B$ translates into a limit on redness for
a given luminosity at a given redshift. In the very worst case 
($z=3$, $M_B=-27.5$), objects redder than $\alpha=-0.7$ are excluded.
On the other hand, by selecting objects bright in 2MASS, we are biased 
against very blue quasars.
There is, however, no statistically significant
correlation between colour and either $M_B$ or $z$; 
and the distribution of spectral indices for the sample is similar to 
(if a little narrower than)
that of the whole LBQS (median $\alpha=-0.3$, Francis et al. 1991).

\subsection{Broad Absorption Line Quasars}
Omont et al. (1996) suggested that those quasars exhibiting broad
absorption lines (BALs) in their optical spectra are preferentially
detected in the submm. This is based on the marginal evidence that
two out of their six $z>4$ IRAM 
detections were BALs, relative to the background 
BAL abundance of $\approx10$\% (Weymann et al., 1991).
We are now in a better position to test this hypothesis.
Considering sources from the current paper,
one of the detections (HS B1141+4201) is unambiguously a BAL, 
another (HS B1310+4308) is possibly a weak BAL. 
In comparison, four of the non-detections are BALs, with a number of others
showing signs of weak absorption, a fraction consistent with the expected
10\%.
Thus, although far from providing proof, these data do not
rule out the Omont et al. hypothesis.
In a forthcoming paper (Priddey et al., in prep.), 
we shall discuss, in detail, 
the evidence from a statistically-significant
combination of all recent submm and mm quasar surveys.

\subsection{Submillimetre versus redshift}
\begin{figure}
\psfig{figure=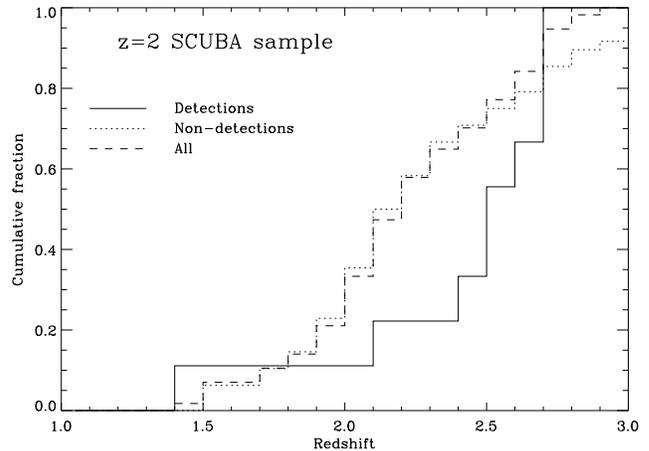,width=90mm}
\caption{Cumulative histogram for redshifts of the $z\approx2$ SCUBA
sample. The difference between the distributions of the detected (solid)
and undetected (dotted) and parent (dashed) samples is clear to the eye,
and supported by formal statistics.}
\label{fig:z2zcumhist}
\end{figure}

\begin{figure}
\psfig{figure=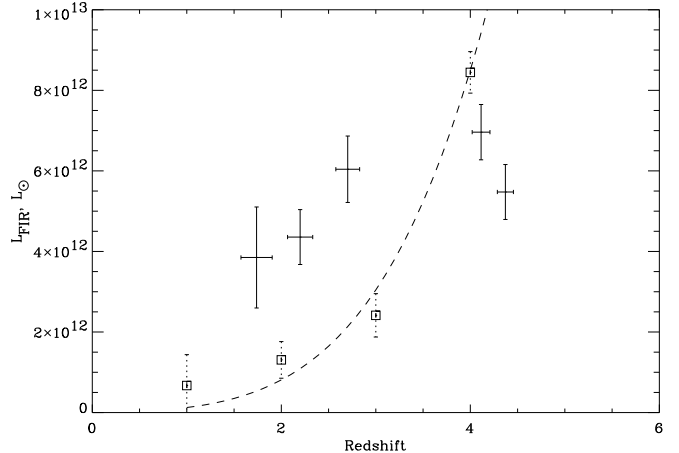,width=90mm}
\caption{FIR luminosity for SBQS quasars (crosses) as a function of redshift, 
determined from the weighted mean of all data within each
bin. Also plotted, for comparison, 
are the radio galaxy points 
(squares) and the power-law fit from Archibald et al. (2001).
}
\label{fig:lfirz}
\end{figure}

Figure \ref{fig:z2zcumhist} is a cumulative histogram of redshift
for the $z\sim2$ sample and for the detected subsample. The 
distributions appear different to the eye, with most of the detections
lying at higher redshift. The median redshift of the sample is $z=2.3$ and 
that of the detections $z=2.6$.
The maximum difference between the distributions of detected and parent 
samples is 0.42, corresponding to a K--S level of significance 
5 percent,
whilst for the the detected and undetected samples, 
the maximum is 0.51, giving a level of significance $\approx$1 percent.
Within this limited redshift range, therefore, there is a suggestion
that the submm detection rate increases with redshift over the range 
$1.5<z<3.0$.
In Figure \ref{fig:lfirz}, the far-infrared luminosity as a function
of redshift has been derived by calculating the $\sigma^{-2}$-weighted
mean of {\it all} the SBQS data--- I02 for $z>3$, and the current work for
$z<3$--- and for detections and non-detections alike.
850$\mu$m flux has been converted to a luminosity assuming
the isothermal SED derived in PM01.
Changing this assumption would shift the $z>4$ and $z\sim2$ points
relative to one another ({\it lowering} the temperature or $\beta$ 
{\it increases} the relative luminosity of the $z>4$ points).
In I02, the fluxes of the non-detections were stacked to give an estimate
of the submm flux of an average $z>4$ quasar---
after checking that sky-subtraction had been performed effectively,
by observing that the distribution of the signal in 
all off-source pixels is a Gaussian with zero mean.
We repeat the experiment with the current $z\sim2$ data, and obtain 
$S=1.9\pm0.4$mJy, which agrees within the uncertainties with the $z>4$ value, 
$2.0\pm0.6$mJy.

We can compare these results with those drawn from
the SCUBA survey of high-redshift radio galaxies (Archibald et al., 2001), 
from which a dramatic increase of FIR luminosity with redshift was inferred,
$L_{\rm FIR}\sim(1+z)^{3-4}$, over the whole redshift range $1<z<4$ 
(see Figure \ref{fig:lfirz}).
While our tentative low-redshift decline echoes this behaviour,
the difference between $z>4$ and $z=2$ is far less marked. 
It is necessary to improve the statistics and to consider the selection
effects before drawing firm conclusions.
Nevertheless, the disparity between the two populations is
a potentially revealing commentary on the nature of radio loudness in AGN. 
Conceivably, radio galaxies follow a more dramatic evolution than
the majority of AGN--- of which they are perhaps the most extreme members---
their radio-loudness originating from a difference in formation mechanism.
Note however that the masses of their central black holes may
be no more extreme than those of the most optically-luminous radio-{\it quiet}
quasars in the present sample, some of which would have $M>10^9\Msun$, 
even assuming accretion at the maximum (Eddington) rate.

The present results show that, on average, the submm properties 
of luminous, radio-quiet quasars at $z\sim2$ are comparable, within the 
uncertainties, to those 
at $z>4$, despite the lack of brighter sources in the present sample.
Although the uncertainties are too great to draw detailed 
conclusions, one can speculate on the physical processes 
underlying the redshift variation of submm detectability at fixed optical
luminosity.
In a starburst scenario, for example, the gas accretion efficiency 
and the star formation efficiency 
could each depend in a different way upon redshift 
(\eg through the dynamical time: Kauffmann \& Haehnelt, 2000).
At $z>4$, the host galaxy is presumably gas- (and dust-) rich, whereas
by $z<2$ most of the gas has formed into stars. 
At yet higher redshifts ($z>5$)
the youth of the universe may preclude the production of enough 
obscuring dust for the quasar to be a luminous submm source
(Priddey et al., in prep.).
Expanding the redshift range of Figure \ref{fig:lfirz} in each direction
is a project currently in progress, but at present it is intriguing
that the average submillimetre loudness seems to decline when the population
itself declines.

\subsection{Caveat: gravitational lensing}
A factor that we have {\it not} addressed in deriving this
result is the influence of gravitational lensing. 
In common with many of bright submillimetre sources found in surveys,
it is likely that fluxes are boosted by lensing. 
This is likely to lead to bias, as the lensing optical depth
increases with redshift
(e.g. Barkana \& Loeb, 2000). Thus the $z>4$ sources are
likely to be intrinsically fainter than they appear, relative to 
those at $z=2$.
Considering individual sources, 
{\it HST} STIS imaging (after the current data were
obtained) showed HS0810+2554 to be lensed into a quadruple source with a 
tight ($<$1 arcsec) separation between the components (Reimers et al. 2001).
This explains the extremely high infrared luminosity implied by its
detection in the {\it IRAS} Faint Source Catalogue.

We stress that {\it no corrections} for lensing have been applied in the
current work, for this is complex and model-dependent.
However, this important issue will be tackled in quantitative detail
in a future paper (Priddey et al., in prep.), where we present a 
detailed statistical analysis of 
all recent JCMT/SCUBA and IRAM/MAMBO high-redshift quasar data.


\section{SUMMARY AND FUTURE WORK}
In this paper, we have presented the first results from a targetted SCUBA
survey of optically-luminous ($M_B<-27.5$), radio-quiet quasars at
$z\sim2$. This is a continuation of the SCUBA Bright Quasar Survey whose
preliminary results, at $z>4$, were reported by Isaak et al. (2002).
The present data confirm the presence of a large scatter in the
correlation between submm and optical luminosity, despite having
minimized the errors in measurement of the latter.
Comparing the $z>4$ and $z<3$ datasets shows that there is no evidence
for a variation of submm properties of luminous quasars across this
redshift range, once one has allowed for a $K$-correction appropriate
for cool, isothermal dust. However, there is a suggestion that
the characteristic submm luminosity increases with redshift between
$z=1.5$ and $z=3$. 
We note that a companion survey, carried out at 1.2mm
with the MAMBO array on the IRAM 30m telescope, reaches similar
conclusions (Omont et al. 2002).

In forthcoming papers, we shall present multiwavelength 
follow-up of bright sources from the present sample, results of 
comparative studies to improve the redshift coverage,
detailed statistical analysis of $>$200 high-redshift
quasars observed at (sub)mm wavelengths,
and astrophysical interpretation of the findings.

\section*{ACKNOWLEDGMENTS}
For support through the period during which the bulk of this work was
carried out, RSP and KGI thank PPARC and RGM thanks the Royal Society.
We are grateful to the JACH staff and those JCMT observers, their projects
displaced by poor weather, who gathered data for us in fallback mode.
The JCMT is operated by JAC, Hilo, on behalf of the parent organisations 
of the Particle Physics and Astronomy Research Council in the UK, the 
National Research Council in Canada and the Scientific Research Organisation 
of the Netherlands.
We thank the anonymous referee for constructive comments.


\end{document}